\documentclass[final]{ws-ijmpd}

\usepackage[T1]{fontenc}
\usepackage[latin9]{inputenc}

\usepackage[super,compress]{cite} 
\usepackage{amsmath}
\usepackage{mathrsfs}
\usepackage{amssymb,amsfonts}
\usepackage{bm,bbm}
\usepackage{graphicx}
\usepackage{color}
\usepackage[usenames,dvipsnames]{xcolor}
\usepackage{hyperref} 
\usepackage[titletoc]{appendix}

\newcommand{\be}{\begin{equation}}
\newcommand{\ee}{\end{equation}}
\newcommand{\ba}{\begin{eqnarray}}
\newcommand{\ea}{\end{eqnarray}}
\def\bs{\begin{subequations}}
\def\es{\end{subequations}}
\def\a{\alpha}
\def\b{\beta}
\def\de{\delta}
\def\g{\gamma}
\def\la{\lambda}
\def\k{\kappa}

\def\Om{\Omega}

\def\G{\Gamma}

\def\s{\sigma}

\def\N{\nabla}

\def\cG{\mathcal{G}}

\def\cI{\mathcal{I}}

\def\cK{\mathcal{K}}

\def\cP{\mathcal{P}}

\def\cV{\mathcal{V}}

\def\ds{d_{\rm S}}

\def\p{\partial}

\def\B{\Box}
\newcommand{\Eq}[1]{(\ref{#1})}
\def\com{\color{magenta}}
\def\cob{\color{blue}}

\newcommand{\au}[2]{#1. #2}
\newcommand{\dat}[1]{, }
\newcommand{\books}[4]{\emph{#1} (#2, #3, #4)}
\newcommand{\oarX}[1]{\href{http://arxiv.org/abs/#1}{{\ttfamily\com arXiv:#1}}}
\newcommand{\arX}[1]{\href{http://arxiv.org/abs/#1}{{\ttfamily\com arXiv:#1}}}
\newcommand{\doin}[6]{\href{http://dx.doi.org/#1}{\cob {\it #2 #3} {\bf #4} (#6) #5}}
\newcommand{\doinn}[5]{\href{http://dx.doi.org/#1}{\cob {\it #2} {\bf #3} (#5) #4}}

\newcommand{\doij}{\doinn}
\newcommand{\tia}[1]{}

\def\rme{\text{e}}
\def\rmd{\text{d}}
\def\rmi{\text{i}}
\def\llangle{\langle\!\langle}
\def\rrangle{\rangle\!\rangle}
\newcommand{\fl}{}

\begin{document}




\title{Quantum spectral dimension in quantum field theory}

\author{Gianluca Calcagni}
\address{Instituto de Estructura de la Materia, CSIC, Serrano 121, 28006 Madrid, Spain\\ calcagni@iem.cfmac.csic.es}

\author{Leonardo Modesto}
\address{Department of Physics and Center for Field Theory and Particle Physics,\\ Fudan University, 200433 Shanghai, China\\ lmodesto@fudan.edu.cn}

\author{Giuseppe Nardelli}
\address{Dipartimento di Matematica e Fisica, Universit\`a Cattolica, via Musei 41, 25121 Brescia, Italy\\
TIFPA -- INFN, c/o  Dipartimento di Fisica, Universit\`a di Trento, 38123 Povo (TN), Italy\\ nardelli@dmf.unicatt.it}

\maketitle

\begin{abstract}
We reinterpret the spectral dimension of spacetimes as the scaling of an effective self-energy transition amplitude in quantum field theory (QFT), when the system is probed at a given resolution. This picture has four main advantages: (a) it dispenses with the usual interpretation (unsatisfactory in covariant approaches) where, instead of a transition amplitude, one has a probability density solving a nonrelativistic diffusion equation in an abstract diffusion time; (b) it solves the problem of negative probabilities known for higher-order and nonlocal dispersion relations in classical and quantum gravity; (c) it clarifies the concept of quantum spectral dimension as opposed to the classical one. We then consider a class of logarithmic dispersion relations associated with quantum particles and show that the spectral dimension $\ds$ of spacetime as felt by these quantum probes can deviate from its classical value, equal to the topological dimension $D$. In particular, in the presence of higher momentum powers it changes with the scale, dropping from $D$ in the infrared (IR) to a value $\ds^{\rm UV}\leq D$ in the ultraviolet (UV). We apply this general result to Stelle theory of renormalizable gravity, which attains the universal value $\ds^{\rm UV}=2$ for any dimension $D$. 
\end{abstract}

\ccode{PACS numbers: 04.60.-m, 11.10.-z, 11.10.Hi}






\section{Introduction and summary of the results}

When geometry becomes quantum, it is common to incur in phenomena of anomalous scaling. An almost universal feature one can find across the most diverse models of quantum gravity is \emph{dimensional flow}, i.e.\ the changing of the dimension of spacetime with the probed scale \cite{tHo93,Car09,fra1}. String theory, asymptotic safety, noncommutative spacetimes, dynamical triangulations, Ho\v{r}ava--Lifshitz gravity, multi-scale spacetimes (by definition), super-renormalizable quantum gravity, black holes, all share this characteristic \cite{AJL4,LaR5,Ben08,ArTr,BeH,MoN,SVW1,SVW2,BMS,Mod1,BGKM,AA,fra6,frc4,frc7,CES,AC1,CM1}. 


\subsection{Spectral dimension and diffusion interpretation}

The operational way to establish dimensional flow is the diffusion equation method. Let us recapitulate the two main ways in which the method is presented in the literature. The first is a naive transposition of the interpretation of the diffusion equation in transport theory. For a classical, nonrelativistic massive particle subject to Brownian motion, the diffusion equation is (see, e.g.\ Ref.\ \citen{frc7} for a derivation)
\be\label{die0}
\left(\frac{\p}{\p t}-\k_1\N_x^2\right)\,P(x,x';t)=0\,,\quad P(x,x';0)=\de(x-x')\,,
\ee
where $t$ is time, $\k_1$ is a constant coefficient, $\N_x^2$ is the Laplacian in $D-1$ spatial dimensions and the delta initial condition encodes the pointwise nature of the object under diffusion. The probability to find the particle at point $x$, after some time $t$ has passed since it started at point $x'$, is the solution $P$ of Eq.\ \Eq{die0}. One can impose the same equation to a spacetime, interpreting the particle as a test probe of the geometry. The variable $t$ is now replaced by an abstract ``diffusion time'' $\s$ and the Laplacian is replaced by the Beltrami--Laplace operator $\B$. Curvature effects are neglected in the calculation, so one can assume the Minkowski metric $\eta={\rm diag}(-,+,\cdots,+)$ and hence $\B=\p_\mu\p^\mu$, where $\mu=0,1,\dots,D-1$. Also, it is part of the definition of the method to Wick rotate the time direction and consider the Euclideanized operator $\B_{\rm E}=\N_x^2$, i.e.\ the $D$-dimensional Laplacian. In effective geometries arising in quantum gravity, the Laplacian is deformed into some more complicated, higher-order or nonlocal operator $\cK$ called \emph{form factor}. In that case, however, the context is field theory, not particle mechanics, and the form factor is defined to be the typical kinetic term of fields. 

This leads to the second formulation of the diffusion equation method, where by ``classical test particle'' one means a real scalar field, which is the field exhibiting the simple kinetic term $-\phi \cK(-\B)\phi$ in the action. $\cK(k^2)=0$ is the relativistic dispersion relation of the field. At this point, to get something like Eq.\ \Eq{die0}, one notes that the propagator of a massless scalar can be written in the Schwinger representation as 
\be\label{Gk}
\tilde G(k^2)=-\frac{1}{\cK(k^2)}=-\int_0^{+\infty}\rmd(\ell^2)\,\rme^{-\ell^2 \cK(k^2)}\,,
\ee
where $\ell$ is a parameter with the engineering dimension of length. This integral representation assumes that $\cK>0$ for all momenta $k$ (thus, it is valid in Euclidean signature). For any $\cK$, the solution of the diffusion equation
\be\label{die}
\left[\frac{\p}{\p\ell^2}+\cK(-\N_x^2)\right]P(x,x';\ell)=0\,,\qquad P(x,x';0)=\de(x-x')
\ee
is the heat kernel 
\be\label{P}
P(x,x';\ell)=\int\frac{\rmd^Dk}{(2\pi)^D}\,\rme^{\rmi k\cdot (x-x')}\rme^{-\ell^2 \cK(k^2)}\,.
\ee
For the integral to converge, one should require $\cK>0$, although this condition does not have to hold for all momenta $k$, contrary to the starting point \Eq{Gk}. Conversely, solutions where $\cK$ can be negative in some regions correspond to physically pathological or unrealistic configurations, as we will find later. Under the assumption (reasonable in the cases of interest) that $k^2=0$ is a pole of $\tilde G$, one has $\cK(0)=0$ and the normalization condition
\be\label{usno}
\langle 1\rangle:=\int\rmd^Dx\,P=1\,.
\ee
The convolution of two $P$'s is also normalized to one, $\int\rmd^Dx\int\rmd^Dx'\,P(x,x';\ell)$ $\times P(x',x'';\ell)=1$. Therefore, if $P>0$ and $\cK(0)=0$,  the solution $P$ can be given  the particle-mechanics interpretation as a probability density function and the running equation \Eq{die} can describe a diffusion process with ``diffusion time'' $\ell$. In a Poincar\'e-invariant spacetime, the return probability is defined as the trace of the heat kernel $P$ over spacetime points:
\ba
\cP(\ell) &:=& \frac{1}{\cV}\!\int\rmd^Dx\,P(x,x;\ell)=\int \frac{\rmd^Dk}{(2\pi)^D}\,\rme^{-\ell^2\cK(k^2)}\nonumber\\
&=&\frac{\Om}{(2\pi)^D}\!\int_0^{+\infty}\rmd k\,k^{D-1}\,\rme^{-\ell^2\cK(k^2)}\,,\label{P1}
\ea
where $\Om$ is a constant coming from the angular integral and $k=|k_{\rm E}|$ is the length of the Euclidean momentum $D$-vector. As customary, one normalizes with respect to a divergent spacetime volume factor $\cV=\int\rmd^Dx$. The \emph{spectral dimension} of spacetime is then defined as
\be\label{ds}
\ds(\ell):=-2\frac{\p\ln\cP(\ell)}{\p\ln\ell^2}\,.
\ee

In the standard free case on Minkowski spacetime, $\cK=-\B$ and the Wick-rotated form factor is simply 
\be\label{fk}
\cK(k^2)=k^2=k_\mu k^\mu=\de_{\mu\nu}k^\mu k^\nu,
\ee
so that the diffusion equation \Eq{die} is
\be\label{die2}
\left(\frac{\p}{\p\ell^2}-\N_x^2\right)\,P(x,x';\ell)=0,\quad P(x,x';0)=\de(x-x').
\ee
The solution \Eq{P} is the Gaussian
\be\label{gau}
P(x,x';\ell)=\frac{\rme^{-\frac{r^2}{4\ell^2}}}{(4\pi\ell^2)^{\frac{D}{2}}}\,,
\ee
where $r^2=|x-x'|^2$. This is positive definite and normalized to one and the probabilistic interpretation holds. The spectral dimension is therefore well-defined and equal, in this case, to $\ds=D$. Effective quantum geometries can induce strong deviations from Eq.\ \Eq{die2} via changes in the Laplacian, the initial condition, the diffusion operator $\p/\p\ell^2$ and the presence of source terms \cite{frc4,CES}. Then, $\ds$ can acquire a nontrivial dependence on $\ell$.


\subsection{Problems with the diffusion interpretation}

The presentation of the spectral dimension in the diffusion interpretation is unsatisfactory for a number of reasons:
\begin{itemize}
\item\emph{Diffusion-time problem.} In the diffusion-equation setting, the parameter $\s\sim\ell$ has no obvious physical meaning {when considered within covariant and background-independent approaches to gravity (either classical or quantum). In fact, in these cases} there is a conceptual as well as operational difficulty to identify a time variable which would lend itself as a ``diffusion time'' of some stochastic process, {unless one fixes the background and makes a choice of coordinates}. On one hand, as discussed below \Eq{die0}, if one defines the diffusion equation with a spatial Laplacian, then there is no justification for assuming Eq.\ \Eq{die0} as it stands, since it is not general covariant and only applies to a nonrelativistic test particle. On the other hand, if one takes the Euclideanized covariant Laplacian, the meaning of $\ell$ and its relation with coordinate time are not obvious. Both impasses are partially solved by regarding \Eq{die} as stemming from the Schwinger representation \Eq{Gk} of the propagator of the probe particle. This explains the form of the diffusion equation, thus cutting the debate between nonrelativistic and relativistic diffusion. However, if the length parameter $\ell$ cannot be interpreted satisfactorily as a diffusion time, what is its physical meaning? In what sense can we talk of a diffusion process in the context of field theory? What does $P$ represent? The spectral dimension is the scaling of $\cP$ with respect to $\ell$, but the absence of a solid physical interpretation of $\ell$ reflects into the lack of one also for $\ds$.
\item\emph{Negative-probabilities problem.} Any theory of quantum gravity with a semiclassical and continuum limit should admit a well-defined profile for the spectral dimension, from the IR down to a (possibly effective) UV scale. In this case, one should give a physical operational meaning to Eq.\ \Eq{die} and $\ds$. Unfortunately, the interpretation of $P$ as a probability density does not hold in some theories, where the heat kernel is not positive definite due to the form the operator $\cK$ takes in such cases. Examples are asymptotic safety and Ho\v{r}ava--Lifshitz gravity \cite{CES}, string field theory and nonlocal gravity \cite{CM1}. These theories are well behaved in any other formal respect, {including unitarity}\footnote{There is still debate about unitarity of asymptotic safety.} (here we are not discussing their phenomenological viability). {For instance, higher-order derivative operators $\B^2$, $\B^3$, \dots lead, in general, to ghosts. Since they are also associated with diffusion equations with no stochastic interpretation (see, e.g.\ Ref.\ \citen{frc4} and references therein), one might be tempted to infer that the negative-probabilities problem is, in theories with such operators, due to the nonunitarity of the theory. However, negative probabilities and unitarity are two independent features. All the examples mentioned above
are either unitary theories or admit unitary formulations \cite{Nie06,PHHN,Mod1}.}\footnote{For some of the examples in the text, there are heuristic reasons why they are unitarity. In Ho\v{r}ava--Lifshitz gravity, higher-order operators contain only spatial derivatives and, therefore, no ghost modes are introduced via Ostrogradski instabilities. In the case of nonlocal gravity and string field theory. entire functions of the form $\rme^\B$ do not introduce extra poles (in particular, instabilities) in the spectrum. In all cases, however, the proof of unitarity is not so intuitive and depends both on the definition of the theory and on detailed quantum calculations.} This means that the problem is not in the internal consistency of these models but in the form of \Eq{die} or, if the form of \Eq{die} cannot be subject to modifications in a given theory, in the diffusion-probabilistic interpretation.
\end{itemize}
These issues pose conceptual problems undermining the physical robustness of the machinery, so often deployed in quantum gravity, leading to the spectral dimension. If there were no operational way to measure $\ds$, it would not make sense as a physical observable describing the geometry of spacetime.

These problems have received scant attention in the written literature. The existence of negative probabilities and possible solutions have been considered in Refs.\ \citen{frc4,frc7,CES}, while the diffusion-time problem has been recognized as an important shortcoming of the definition of $\ds$ in Refs.\ \citen{AAGM1} and \citen{AAGM2} (where it was even proposed to abandon $\ds$ as a meaningful observable). Outside these examples, the general tendency, which we presently discourage, is to ignore the above points, to trust the number $\ds$ at face value and regard its definition as purely mathematical. 


\subsection{Aim of the paper and main results}

In this paper, we attempt to solve these issues within the frame of quantum field theory. {Under certain but quite general assumptions, one can construct a physically meaningful definition of $\ds$. These assumptions are:
\begin{itemize}
\item The theory lives on a continuum and admits a Minkowski background. Models of quantum gravity based on discrete structures can be included if they admit a continuum limit where an effective propagator encoding some generic dispersion relation $\cK(k^2)=0$ can be formulated. Lorentz invariance is not required but, for the sake of simplicity, we will present all the formul\ae\ in a Lorentz-invariant setting. For the QFT interpretation to be valid in a regime where an effective field theory can be formulated, both this regime and dimensional flow should occur at scales larger than the characteristic scale $L$ of the fundamental degrees of freedom of the putative quantum-gravity theory. While \emph{a priori} this expectation seems hard to be fulfilled in general, it is actually corroborated by concrete and highly nontrivial examples. In loop quantum gravity and group field theory, dimensional flow takes place at scales $\ell\gg L$ larger than those of the underlying combinatorial discrete structure \cite{COT2,COT3}. Another instance is causal dynamical triangulations: although the phase space of the theory has different regions with properties even wildly different from ordinary geometries, dimensional flow to the IR occurs in the so-called phase C, where geometry is approximated by a de Sitter spacetime at scales much larger than the average size of the triangulation cell \cite{AJL4}. Thus, even if the assumption of continuity seems too strong and bound to be violated in quantum gravity at large, top-bottom approaches indicate otherwise. We are not aware of any counterexample.
\item The theory admits a well-defined analytic continuation to Euclidean spacetime. In the diffusion interpretation, Wick rotation is not necessary if one is only interested in the spectral dimension of space alone: in that case, one defines a diffusion equation on spatial slices and limits the analysis to that context, without the need to Euclideanize. This is not possible in the QFT interpretation, since time and space are entangled relativistically in the propagator. This is not an issue, however, since by definition dimension observables of a space\emph{time} (spectral, Hausdorff and walk dimension) are always constructed in the Euclidean setting. On the other hand, all the major quantum-gravity examples we mentioned (asymptotic safety, Ho\v{r}ava--Lifshitz gravity, nonlocal gravity and string field theory) and the one we will  construct (Stelle theory) obey the Osterwalder--Schrader conditions \cite{OsSc1,OsSc2} in their Euclidean formulation, which guarantee the validity of the analytic continuation. 
\item The propagator has a pole at $k^2=0$, so that $\cK(0)=0$. As we already said, physically meaningful cases have $\cK(k^2)\geq 0$ but our treatment will allow us to study also instances where $\cK$ becomes negative in a certain range of the momenta.
\item $P(x,x';\ell)$ and its massive version $P_m(x,x';\ell)$ are nonsingular at $x=x'$ for any positive value of the resolution scale, $\ell>0$. This condition, respected in almost all cases in quantum gravity (see Refs.\ \citen{frc4,frc7,CM1,CES} for examples), is necessary to have a well-defined spectral dimension at infinite resolution.\footnote{The only instance we are aware of where this condition is not satisfied is the effective nonlocal model of Ref.\ \citen{AC1}. There, the diffusion process is well-defined only above a certain minimal diffusion length (or resolution scale) $\ell_*$ of order of the Planck scale. To include also this and similar cases in our treatment, it would be sufficient to require that $P(x,x';\ell)$ is nonsingular at $x=x'$ for any value of the resolution scale above a certain critical value $\ell>\ell_*\geq 0$. Then, all the $\ell$ integrals in our paper should be performed in the range $[\ell_*,+\infty)$ rather than $[0,+\infty)$. We do not attempt this generalization here.}
\end{itemize}
}
When the probe is quantum, the differential operator $\cK(-\N^2)$ is interpreted as stemming from the quantum propagator of the particle field, computed at one- or higher-loop level in ordinary perturbation theory ($\cK(k^2)\neq k^2$). In Sec.\ \ref{4}, we shall focus on a class of logarithmic dispersion relations:
\be\label{ef}
\cK(k^2)=k^2\left[1+a\, k^{2n}\ln \left(\frac{k^2}{E^2}\right)\right]\,,
\ee
where $a$ is a constant with momentum scaling $[a]=-2n$, $n\in\mathbb{N}$ and $E$ is a reference energy scale. The form of $\cK$ is sufficiently general to include most of the (one- and two-loop resummed) renormalizable massless field theories.\footnote{Both $\lambda \phi^4$ in $D=4$ dimensions and $\lambda \phi^3$ in $D=6$ have a loop-corrected inverse propagator which is precisely of the type \Eq{ef}, although $a<0$ in those cases (see, for instance, Refs.\ \citen{Pes95,Ram01,Sre07}). More generally, logarithmic form factors sometimes appear in quantum gravity, as in proximity of a black hole obeying the entropy-area law \cite{AC1,Pad98} or in $\kappa$-Minkowski noncommutative spacetime \cite{AM}, but these specific cases have some subtle differences with respect to \Eq{ef} which would require a separate discussion (see the previous footnote).}

To summarize our main results, we reinterpret all the details of the definition of the spectral dimension within quantum field theory. The interpretation problem is clarified by regarding $\ell$ as a resolution scale and the diffusion equation as a probing of the system under a resolution-varying tool (Sec.\ \ref{2}). The role of the mass of the particle probe is discussed in Appendix \ref{A0}.

The resolution interpretation, partly formulated in Ref.\ \citen{CM1} but sometimes invoked also in some of prior literature, regards $\ell$ as a resolution scale and \Eq{die} as a running equation. It solves the issues connected with the diffusion picture in covariant theories, but it still regards $P$ as a classical probability. A step further in solving the negative-probabilities problem consists in abandoning also this last inheritance from the diffusion-equation interpretation and regard $P$ as a \emph{quantum probability amplitude} rather than a classical probability (Sec.\ \ref{3}). By this assumption, and recalling that the heat kernel is related to a quantum propagator, the return probability is reinterpreted as the effective contribution of the loop-corrected vacuum-to-vacuum amplitude of the probe field \emph{at a given resolution}, and the quantum spectral dimension is nothing but the scaling of this contribution. A regulator is introduced at this point to render bubble diagrams and several formal expressions finite. 

It is important to stress once again the practical usefulness of the new interpretation of the spectral dimension. Even if the number $\ds$ for a given theory is not changed (and thus all previous quantitative results in the literature remain valid), its derivation is put on a ground which does not suffer from the issues of the fictitious diffusion process assumed in older treatments. The disappearance of diffusion time ``rehabilitates'' the physical meaning of the spectral dimension of space\emph{time}, not only in covariant theories but also in cases (such as Ho\v{r}ava--Lifshitz gravity) where time plays a separate role. Moreover, the theories where a negative probability arises in the diffusion interpretation \cite{CES,CM1} benefit from the QFT reinterpretation of what the actual probability should be.\footnote{Another possible advantage of the QFT interpretation is that we give an operational definition of the spectral dimension. Previously, there were only vague ideas about ``placing a test particle in a spacetime and see how it diffuses.'' Beyond this wishful thinking, the only quantitative exploration of the ``observable'' consequences of such picture was given in Ref.\ \citen{frc7}, where the actual random walks of the test particle in some generic anomalous spacetimes (possibly of quantum-gravity origin) were plotted. On the other hand, the QFT interpretation of the present paper may open up new possibilities towards an actual measurement of the spectral dimension. The abstract ``diffusion time'' takes the role of the energy scale at which a particle experiment is done. $\ds$ is tightly related to the vacuum diagrams of a QFT and, as we will see, different particles can see, surprisingly, different spectral dimensions. These pieces of information are still fragmentary but they might serve to create a connection between $\ds$ and experiments.}

The general setting we shall provide is quite flexible and the results of Sec.\ \ref{4} will be valid for any QFT with inverse propagator of the form \Eq{ef} and respecting the assumptions stated above. In particular, a dispersion relation of the form \Eq{ef} is that of the one-loop graviton propagator in renormalizable Stelle theory \cite{Ste77,Ste78,FrZ82}. In this case, the probe is the graviton, i.e.\ a quantum fluctuation of spacetime, so quantum effects can be ascribed to geometry itself. In Sec.\ \ref{stel}, we will calculate the spectral dimension of Stelle gravity for the first time.


\section{Interpretation problem}\label{2}

According to the resolution interpretation \cite{CM1}, one places a test particle on the geometry one wants to probe at some initial point $x'$ (suitably defined also in discrete geometries), and asks what the probability is to find the particle in a neighborhood of size $\ell$ of another point $x$ when the system is probed with a resolution $1/\ell$. The length scale $\ell$ in the Schwinger propagator \Eq{Gk} (or \Eq{Gkm}, for a massive field; this case is considered in Appendix \ref{A0}) represents the minimal detectable separation between points. In particle-physics experiments, $1/\ell$ can be identified with the energy scale at which scattering processes take place. Note that here we are taking two independent steps: one is to stress the origin of the diffusion equation from the Schwinger representation of the propagator and the other is to embrace the resolution interpretation (\emph{a priori} and by itself, the integration parameter in the Schwinger representation is no more physical than an abstract diffusion time).

From Eq.\ \Eq{Gk}, we can see that the heat kernel $P(x,x';\ell)$ is the Green function at a given resolution:
\be\label{Gk2}
G(x-x')=-\int_0^{+\infty}\rmd(\ell^2)\,P(x,x';\ell)\,,
\ee
where the dimensionality of these quantities in momentum units is $[G]=D-2$, $[\ell]=-1$ and $[P]=D$. 
 Consider a generic form factor $\cK(k^2)$. In QFT, loop corrections modify the propagator of a scalar particle and, consequently, $\cK(k^2)\neq k^2$; in a Lorentz-invariant setting, $\cK$ will depend on the square $k^2$ of the momentum $D$-vector.

We now show that, if the running equation \Eq{die} holds, then $G$ respects the Green equation. In the following, we will adopt the Euclidean version of a Minkowski background. For a massive Green function, one has 
\ba
\de(x-x') &=& -\cK(-\N_x^2)\,G(x-x') = \int_0^{+\infty}\rmd(\ell^2)\,\cK(-\N_x^2)\,P(x,x';\ell)\nonumber\\
      		& \stackrel{\text{\tiny \Eq{die}}}{=} & -\int_0^{+\infty}\rmd(\ell^2)\,\p_{\ell^2} P(x,x';\ell) = P(x,x';0)-P(x,x';+\infty)\,,\label{gp}
\ea
and we recover the ``initial condition'' (shape of the probe at infinite resolution) if $0=P(x,x';+\infty)$, which is verified in all the field theories (including of quantum gravity) that are the context of the present paper.\footnote{This is obvious from the physical interpretation of the condition $P(x,x';+\infty)=0$ in either the diffusion or resolution-based picture: the probability to find the particle at point $x$ when probing spacetime in the limit of zero resolution ($1/\ell\to1/\infty$) is zero. This property can also be checked explicitly by looking at the expression of the heat kernel in the quantum-gravity models under consideration: see Ref.\ \citen{CES} (Eqs.\ (9) and (10)) for asymptotic safety and Ho\v{r}ava--Lifshitz gravity, Ref.\ \citen{CM1} (Eq.\ (27)) for string field theory and nonlocal theories of gravity, Refs.\ \citen{frc4} and \citen{frc7} for other heat kernels (in Eq.\ (74) of Ref.\ \citen{frc4}, $s$ should be $\sigma$) and Appendix \ref{A1} of the present paper for Stelle's gravity. In all these cases, $P$ can be written as $P\sim f_1(\ell) f_2[(x-x')^2/\ell]$ either exactly or in the large-$\ell$ limit. But $f_2$ is finite when sending $\ell$ to infinity (otherwise, one could not define the trace of $P(x,x;\ell)$) and the remainder $f_1(\ell)$ typically goes to zero (as an inverse power law $\sim\ell^{-D}$, asymptotically) when $\ell$ diverges.} Note also that \Eq{gp} is the precise expression of possible divergences of \Eq{Gk2}. It states that the integral of $P$ in $\ell$ is singular at $x=x'$, even if $P$ itself is not (a very general assumption we made in the introduction).


\section{Negative-probabilities problem}\label{3}

In the previous section, we have given the running equation \Eq{die} an operational meaning in QFT. Before computing the spectral dimension according to the scheme of the introduction, one must ensure that it is physically well-defined. In the diffusion interpretation, the solution $P$ is a classical probability density normalized to 1. In particular, $P$ must be positive semi-definite. However, in many situations this is not the case, the solution of the diffusion equation is negative for certain values of diffusion time and spacetime points $x$, and there is no diffusion interpretation at all, even if the ``return probability'' $\cP$ is positive-definite \cite{frc4,CES} (we use quotation marks since, in this case, it is a misnomer). In other words, there is no diffusing process by which the test particle can be found ``somewhere'' after a given diffusion time and, consequently, no clear geometric and physical meaning can be attached to the quantity \Eq{ds}. Notably, the polynomial form factor \Eq{barf} falls into this pathological class of models when $n\geq 1$ \cite{frc4,CES}. We numerically checked that the same behaviour occurs for Eq.\ \Eq{ef} as well as for the exponential operator $\cK(k^2)=k^2\rme^{\ell_{\rm s}^2k^2}$ typical of string field theory, where $\ell_{\rm s}^2\propto \a'$ is proportional to the Regge slope. In all these cases, $P<0$ for some values of $x-x'$ and $\ell$.

The problem persists in the resolution-based picture, since its main difference with respect to the diffusion interpretation is in the interpretation of $\ell$, while $P$ is still the probability density associated with finding the probe at some point on the effective manifold. In a yet alternative revisitation of the problem, the effect of nonstandard dispersion relations can be mapped into a nontrivial measure of momentum space, so that the spectral dimension is identified with the Hausdorff dimension of momentum space, at least in the UV limit \cite{AAGM1,AAGM2}. However, one still lacks an interpretation for the function $P$, Eq.\ \Eq{P}, from where $\ds$ is derived.

Adopting a QFT interpretation of the running equation \Eq{die} can do the job. From Eq.\ \Eq{Gk2}, we see that the heat kernel is related to a transition amplitude, which suggests to give $|G|^2$ (not $P$) the meaning of a probability density function in the sense of a QFT path integral. This is wholly different from the usual probabilistic interpretation, where the diffusion equation can be derived from stochastic quantum mechanics and $P\sim \llangle |\cG^{\rm R}(x,t;x',t')|^2\rrangle$ is actually the \emph{square} of a mechanical-particle retarded propagator, averaged over the noise \cite{frc7,AkM}. While the quantum-mechanics interpretation is helpful to derive (rather than assuming) the diffusion equations \Eq{die} or \Eq{die2} from fundamental (albeit not quite first) principles \cite{frc7}, the QFT interpretation of $P$ as a probability amplitude rather than a probability density is perhaps more natural when one has to calculate the heat kernel of an effective spacetime. The key difference between the two pictures is the role of diffusion time: in the QFT case, it is a resolution scale, an arbitrary parameter.

Therefore, {we set the probability for a particle field to propagate from $x'$ to $x$ to be proportional\footnote{In general, the proportionality constant may be infinite. Divergences like this can be parametrized by regularization procedures such as the one discussed below in the text.} to} the squared transition amplitude (integration domain $\ell\in[0,+\infty)$ omitted here and below)
\ba
\fl |G(x-x')|^2 &=&\!\! \int\rmd(\ell^2)\,\rmd({\ell'}^2)\,P(x,x';\ell) P^*(x,x';\ell')\label{g2}\\
						 &=&\!\! \int\rmd(\ell^2)\,\rmd({\ell'}^2)\int\frac{\rmd^Dk}{(2\pi)^D}\frac{\rmd^Dk'}{(2\pi)^D}\,\rme^{\rmi (k-k')\cdot (x-x')}\rme^{-\ell^2 \cK(k^2)-{\ell'}^2 \cK({k'}^2)}.\nonumber
\ea
In the present reinterpretation of the spectral dimension, the diffusion of a particle under a classical random stochastic process is replaced by its quantum propagation from one point to another. We can also find a relation between the return probability $\cP$ (that indeed has a probabilistic interpretation) and the square of the absolute value of the Green function. Integrating \Eq{g2} in $x$,
\ba
\fl\int\rmd^Dx\,|G(x-x')|^2 &\stackrel{\text{\tiny \Eq{g2}}}{=}& \int\rmd({\ell''}^2)\,\rmd({\ell'}^2)\int\frac{\rmd^Dk}{(2\pi)^D}\,\rme^{-({\ell''}^2+{\ell'}^2)\cK(k^2)}\nonumber\\
\fl						             &=& \int\rmd({\ell''}^2)\,\rmd({\ell'}^2)\,\cP(\sqrt{{\ell''}^2+{\ell'}^2}) \nonumber\\
\fl                      &=& \int_0^{+\infty}\!\rmd({\ell''}^2)\int_{{\ell''}^2}^\infty\!\rmd(\ell^2)\,\cP(\ell)\nonumber\\
\fl                      &=&  \int_0^{+\infty}\!\rmd(\ell^2)\int_0^{\ell^2}\!\rmd({\ell''}^2)\,\cP(\ell)\nonumber\\
\fl						             &=& \int_0^{+\infty} \rmd({\ell}^2)\, {\ell}^2\,\cP(\ell)\,,\label{fin1}
\ea
where we made the change of variables $\ell=\sqrt{{\ell''}^2+{\ell'}^2}$. 

Equations \Eq{g2} and \Eq{fin1} deserve several comments. According to our hypothesis, $P$ is a smooth function (verified in the great majority of quantum-gravity models and, in general, in QFT) but its double integral in \Eq{g2} diverges at least at the lower boundary $\ell=0$. The mathematical reason is obvious: in the language of the diffusion interpretation, this lower boundary is the initial condition for the diffusion process, where by definition the heat kernel is a delta (i.e.\ the diffusion probe is pointwise). We can see this in the cornerstone example of the Gaussian heat kernel \Eq{gau}. The associated return probability is the inverse power $\ell^{-D}$, so that the integral of $\ell^{2-D}$ diverges at $\ell=0$ for positive $D$. This is consistent with the left-hand side of \Eq{fin1}. In $D$ topological dimensions, the Euclidean Green function is $G(r)\propto r^{2-D}$ and its square is $|G(r)|^2\propto r^{2(2-D)}$. This distribution is well-defined everywhere for $r>0$ but it is singular in $r=0$, which is the only singular point of the heat kernel when $\ell=0$. So, when one integrates $|G|^2$ in $x$, one has a left-hand side of the form $\sim\int_0^\infty \rmd r\,r^{D-1}r^{2(2-D)}=\int_0^\infty \rmd r\,r^{3-D}$. In $D=4$, this integral also has an IR divergence, but introducing a mass removes this problem. (Incidentally, this is yet another interesting application of the need for a nontrivial mass in the QFT interpretation.) The UV divergence cannot be removed and expressions such as \Eq{g2} and \Eq{fin1} are actually formal. 

One way to remove the UV divergence is to regularize Eq.\ \Eq{g2}. Dimensional regularization would clearly not work (by definition, the dimensional regularization of the integral of a power law is zero). The alternative is to introduce a UV cutoff $\bar\ell>0$ and perform the integrals in \Eq{g2} and \Eq{fin1} with this cutoff. In a sense, this would correspond to avoid the dangerous point $x\sim x'$. This cutoff can have a physical justification: an infinite resolution would be in conflict with the general principles of quantum mechanics (Heisenberg principle). On the other hand, the present discussion is focused on the operational aspects of the spectral dimension and, clearly, any experiment will have a finite resolution (characteristic energy scale).

To avoid confusion here, it is important to recall that the sole purpose of Eqs.~\Eq{g2} and \Eq{fin1} is to answer the conceptual questions: 1) If we give up the diffusion interpretation and adopt the QFT interpretation, then we have to give up also the interpretation of the heat kernel as the probability that the particle propagates from $x'$ to $x$. So what is such probability in the QFT interpretation? 2) The spectral dimension is the scaling of the return probability, but what is the return probability in the QFT interpretation? The mathematical definition of the spectral dimension $\ds$ is untouched, one does consider the full range of $\ell$ as well as the point $x=x'$ (otherwise, it would be impossible to take the trace). However, Eqs.\ \Eq{g2} and \Eq{fin1} are intrinsically divergent. To make these equations mathematically meaningful, one has to introduce a cutoff $\bar\ell$, in which case both the left- and right-hand side of Eq.~\Eq{fin1} are positive semi-definite. In any other respect, our discussion on the spectral dimension is unchanged. In particular, the regulator $\bar\ell$ does \emph{not} affect dimensional flow in any appreciable way, for two reasons. First, $\bar\ell$ is used to make sense of the formal expressions \Eq{g2} and \Eq{fin1}, which are helpful to complete the QFT interpretation of the spectral dimension but neither of which are ever used in the calculation of the $\ds$. Second, as stated in the introduction, continuous dimensional flow usually happens at scales larger than the characteristic scales of the fundamental quantum-gravity degrees of freedom. Thus, the presence of a regulator does not interfere with the generation of a dimensional flow by the building blocks at ultra-microscopic scales and it is possible to take $L<\bar\ell\ll\ell$ in the effective QFT regime without any inconsistency.

Another interesting consequence of the QFT picture is the physical interpretation of $\ds$. The return probability summed over all scales $\ell$ turns out to be the contribution of vacuum-to-vacuum diagrams, i.e.\ the integration of a loop-corrected propagator at a given resolution $1/\ell$. From Eq.\ \Eq{Gk2},
\be\label{P3}
-\frac{1}{\cV}\int\rmd^Dx\, G(0)=-G(0)=\int_0^{+\infty}\rmd(\ell^2)\,\cP(\ell)\,.
\ee
If $G$ is the tree-level propagator (Eq.\ \Eq{fk}), this is the one-loop bubble diagram. If $G$ is the loop-corrected propagator, this expression is an effective contribution. Thus, the spectral dimension \Eq{ds} is the $\ell$-scaling of this contribution for a given resolution in the massless limit. In the QFT interpretation, the left-hand side of Eq.\ \Eq{P3} represents the contribution of a loop diagram with no external legs, i.e.\ a bubble contribution to the vacuum energy of the particle. If the theory is perturbatively renormalizable, then these diagrams should be renormalized order by order in the loop expansion.

These simple relations between Green function and $\cP$ stem from the fact that the Fourier transform of ``$|P|^2$'' is additive in $\ell^2$. The focal object to pay attention to is therefore the return probability $\cP$, not $P$. This is often done in the literature, but here, we provided an explicit justification for it. Positivity of $\cP (\ell)$ is satisfied in all known examples of quantum-gravity and string-field-theory dispersion relations. In particular, this result finds an immediate application to asymptotic safety and Ho\v{r}ava--Lifshitz gravity. In those cases, a solution was proposed where the diffusion equation \Eq{die2} received special modifications \cite{CES}. In asymptotic safety, these modifications stemmed from novel cutoff identifications, while in Ho\v{r}ava--Lifshitz gravity source terms were added by hand in the diffusion equation. However, the physical origin of such source terms is not clear. Here we offer the QFT interpretation as a portable alternative, potentially valid in all QFT approaches to quantum gravity such as those just mentioned and Stelle gravity (see Sec.\ \ref{stel}).

\section{Quantum spectral dimension}\label{4}

The diffusion-probabilistic interpretation of the spectral dimension does not provide a clear way to actually measure this number. Beside the conceptual issues discussed above, it is difficult to imagine an experiment at high energies or ultra-high curvature where an abstract pointwise probe is let diffuse via a stochastic process ideally generated \cite{frc7,CES} by the pushing-around of quanta of geometry. On the other hand, the QFT interpretation can provide a more realistic setting where the probe is quantum and where one can hope to make actual measurements with particle-physics experiments. To test the most immediate consequences of this possibility, in this section we abandon the remote realm of quantum gravity and consider the more conventional setting of QFT on flat spacetime. 

In this case, the probe (i.e.\ the particle field involved in the experiment) propagates on a fixed background: spacetime is not dynamical and, therefore, there are no back-reaction effects. Thus, one takes a specific field (with its characteristic kinetic operator) and asks how spacetime dimensionality is felt by such field. From this exercise, we learn that, contrary to the case of an abstract pointwise generic ``probe,'' different quantum particles can perceive different spacetime geometries. We identify the source of the effect in the way the form factor $\mathcal{K}$ is modified at the quantum level. The fact that not only Minkowski spacetime is not seen as $D$ dimensional at all scales but also that different quantum particles can see different effective dimensions in the UV is therefore not so surprising: perturbative corrections to the effective kinetic term can vary according to the type of field considered. This suggests, as advertized in the introduction, to introduce the concept of ``quantum spectral dimension,'' as opposed to the traditional or ``classical'' one.\footnote{For a different application of this terminology in discrete quantum-gravity settings, see Refs.\ \citen{COT2} and \citen{COT3}.}

In the Appendices, we estimate analytically the trace of the heat kernel \Eq{P1} for the function \Eq{ef} when $a>0$ in two different regimes: the IR limit (small resolution, $\ell E\gg 1$) in Appendix \ref{A1} and the UV limit (large resolution, $\ell E\ll 1$) in Appendix \ref{A2}. {As we will see, a change of spacetime dimensionality is mainly due to the classical higher-order power $(k^2)^n$ rather than the logarithm. In this respect, it is not at all surprising to have dimensional flow with the form factor \Eq{ef}, which is similar to that for the polynomial form factor (e.g.\ Refs.\ \citen{SVW2,frc4} and references therein; this is also the form factor of Stelle's theory at the tree level)
\be\label{barf}
\bar \cK(k^2)=k^2[1+a\, k^{2n}]\,.
\ee
Still, the analytic formul\ae\ we shall obtain will highlight some interesting properties of Eq.\ \Eq{ef}. For instance, the UV or high-resolution limit of $\ds$ is independent of the parameters $a$ and $E$, a fact that will allow us to comment also on cases (mainly, scalar field theories) where $a<0$. Also the IR or low-resolution limit is nontrivial. In fact, the magnitude of $a$ does matter and there is a critical value $a_*(E,n)$ above which the result $\ds^{\rm IR}=D$ is \emph{not} recovered. The practical reason is that, for $a>a_*$, new poles appear in the propagator and the theory becomes unphysical. To keep the same particle spectrum at any order in perturbation theory, any well-defined quantum field theory must predict a parameter $a$ below this critical value, which is the case for all models considered here. No such threshold exists for models with dispersion relations of the form \Eq{barf}.}


\subsection{Small-resolution (IR) limit}\label{ir}

Against intuition, the analysis of the IR limit is not trivial. One would expect that, by removing the cutoff scale $E$, one would obtain immediately $\cP\sim \ell^{-D}$, from which the small-resolution limit
\be\label{dsir}
\ds\stackrel{\text{\tiny IR}}{\simeq} D
\ee
descends. However, this is true only within a certain region in the parameter space and not for all values of the dimensionless constant 
\be
\a:=a E^{2n}\,.
\ee
From the analysis given in Appendix \ref{A1}, it turns out that the behaviour of the return probability drastically changes at the critical value
\be\label{a*}
\a_{*}:=n\,\rme\,,
\ee
where $n$ is the integer in Eq.\ \Eq{ef}. Up to a positive normalization constant, from Eqs.\ \Eq{Pya} and \Eq{Pyb} one has
\be\label{Pl1}
\cP(\ell)\ \stackrel{\ell E\gg 1}{\propto}\ (\ell E)^{-D}\qquad \text{($0<\a<\a_{*}$ with $n\neq 0$)}\,,
\ee
while Eqs.\ \Eq{tiI} and \Eq{Pycd} yield
\be\label{Pl2}
\cP(\ell)\ \stackrel{\ell E\gg 1}{\propto}\ \rme^{-(\ell E)^2 \cK(k_{\rm min}^2)}\sqrt{\frac{2\pi}{(\ell E)^2\ \cK''(k_{\rm min}^2)}}\qquad \text{($\a>\a_{*}$, $n\geq 0$)}\,,
\ee
where double primes denote the second derivative with respect to the argument, and
\ba
&&k_{\rm min}(n\neq 0) = E\left\{-\frac{n}{\a(n+1)\,W_0\left[-\frac{n\rme^{n/(n+1)}}{\a(n+1)}\right]}\right\}^{\frac{1}{2n}},\\
&&k_{\rm min}(n=0) = E\,\rme^{-\frac{1+\a}{2\a}}
\ea
is the momentum value at which the form factor \Eq{ef} acquires the relative minimum. If $\a > \a_{*}$, this is also the absolute minimum and $\cK(k_{\rm min}^2)<0$.
In the above equation, $W_0$ is the principal branch of Lambert's $W$ function (see Appendix \ref{A1} for details).

Thus, Eq.\ \Eq{dsir} is valid only in the parameter range $0<\a<n\,\rme$ (which is nothing but the condition $\cK>0$), while in the other case $\ds\simeq 2(\ell E)^2\cK(k_{\rm min}^2)\to-\infty$. What happened? The critical value \Eq{a*} signals the appearance of extra poles in the propagator (zeros of $\cK$), as one can see from cases (c) and (d) in Fig.\ \ref{fig3}. Such a change in the particle spectrum should not occur in any sensible quantum field theory and, moreover, it could be incompatible with the renormalization scheme adopted. Also, for $\a>\a_*$ the form factor $\cK$ is negative definite and Eq.\ \Eq{Gk} is invalid. We can conclude that the limit \Eq{Pl2} is either unphysical or, at best, it does not describe the theory we started from. This problem does not arise with the form factor \Eq{barf}, which is positive definite for $a>0$ and there is no upper bound for $\a$.


\subsection{High-resolution (UV) limit}\label{uv}

In the deep UV, the spectral dimension of spacetime as seen by the propagating particle drops down to a universal value independent of the parameters $\a$ and $n$. From Eq.\ \Eq{ciuv}, it follows that
\be\label{Puv}
\cP\ \stackrel{\ell E\ll 1}{\propto} \left[-\a(\ell E)^2\ln(\ell E)\right]^{-\frac{D}{2(n+1)}},
\ee
so that
\be\label{dsuv}
\ds\ \stackrel{\text{\tiny UV}}{\simeq}\ \frac{D[1+\ln (\ell E)^2]}{(n+1)\ln (\ell E)^2}\to\frac{D}{n+1}.
\ee

The ordinary QFT of a scalar field on Minkowski spacetime is somewhat special. In fact, $n=0$ and the value of the constant $a$ in Eq.\ \Eq{ef} is negative, both at the one-loop level in the $\lambda \phi^3$ model in $D=6$ and at the two-loop level for the massless $\la\phi^4$ theory in $D=4$ \cite{Sre07}. These cases are problematic, as $a<0$ makes the integral \Eq{P1} divergent and prevents the validity of the asymptotic behaviours given in Appendices \ref{A1} and \ref{A2}. Even invoking analytic continuation, difficulties remain on the physical side: the return probability \Eq{Puv} is negative definite if $\a<0$, and there is no probabilistic interpretation for the diffusing process. On the other hand, the final result \Eq{dsuv} does not depend on $a$, which suggests to take the limit $n\to 0$ and argue that form factors \Eq{ef} with $n=0$ do not trigger a running in the spectral dimension, which remains at its classical value $\ds=D$. This is compatible with the well-known notion that the simplest occurrence of dimensional flow, when no other ingredient of the theory is modified, takes place in the presence of polynomial dispersion relations $\cK(k^2)=k^2(1+k^{2n}+\dots)$, i.e.\ in models with derivatives higher than (or, more generally, different from) second-order. Under this perspective, higher momenta (higher curvature terms in Stelle quantum gravity) are responsible for the feeling of different spectral dimensions at different scales, while the quantum nature of the probe itself is not enough to introduce anomalous scalings. In fact, unless one chooses exotic scalar field theories with higher-order or fractional derivatives, all the standard renormalizable scalar theories have $n=0$ in Eq.\ \Eq{ef}, as it follows by a simple dimensional analysis. Therefore, we may conjecture that the right-hand side of \Eq{dsuv} would always be $D$ for a standard quantum scalar field, which would not experience any difference in the UV and IR spectral dimensions. The case of Stelle gravity considered in the next section is more interesting and does not suffer from these problems, since $a>0$.
 

\section{Stelle quantum gravity}\label{stel}

On a discrete structure, genuine quantum-gravity effects can arise from the choice of states in the quantum theory, such as in causal dynamical triangulations \cite{AJL4,BeH} or loop quantum gravity and spin foams \cite{COT2,COT3}. On a continuum, they typically arise from modifications (differently motivated in each case) of the Laplacian (as in asymptotic safety \cite{LaR5,CES}, Ho\v{r}ava--Lifshitz gravity \cite{CES,Hor3}, nonlocal gravity \cite{Mod1,CM1} and noncommutative spacetimes \cite{ArTr,AA}), or from a change in the differential structure of geometry itself as in multi-scale spacetimes \cite{frc7}. Both factors are equally important and intertwined in multi-scale spacetimes as well as in the discrete approach of causal sets \cite{EiMi}. Even at the classical level of both probe and geometry, the spectral dimension does not coincide with the topological one whenever curvature bends the background. In all the above cases and whenever a continuum limit can be identified, the spectral dimension is usually defined to probe effects of the quantum geometry in the spectrum of the Laplacian, while at the same time excluding trivial effects due to curved backgrounds. Therefore, in quantum gravity it is part of the conventional definition of $\ds$ to consider propagation of the probe on a flat space or its discrete counterpart. Not doing so would otherwise result in the impossibility to disentangle spurious curvature effects from phenomena beyond classical gravity.

A concrete realization of the form factor \Eq{ef} is provided by Stelle's quantum gravity, a four-dimensional higher-derivative theory where gravity is quantized perturbatively \cite{Ste77,BOS}. This is not a viable model of quantum gravity because it contains ghost modes, but it provides an easy application to quantum gravity of the interpretation and method presented in this paper. Also, it is of academic interest to determine the spectral dimension of this theory, which is older by more than a decade than the models presently under scrutiny in the community. To get information of the quantum geometry and determine the spectral dimension, we will apply the philosophy of Sec.\ \ref{4} to graviton fluctuations on flat spacetime {(which are well defined, the theory being quantized perturbatively)}, when quantum corrections are included in the propagator. These quantum corrections will drive a nontrivial running of $\ds$. Note that, as in any other approach to quantum gravity on a continuum, we interpret the running of the spectral dimension as due to quantum geometry, even if the background around which we consider the fluctuations is (by necessity, according to what said in the previous paragraph) flat spacetime. In fact, the profile of $\ds$ is mainly determined by the form of the kinetic operator of the probing particle and such operator, in turn, determines the renormalization-group properties of the theory.

The generalization of Stelle's theory in a $D$-dimensional spacetime reads 
\be
S =\int \rmd^D x \sqrt{-g} \left[\frac{2}{\kappa_D^{2}} R + \frac{\g}{2} R^2 + \frac{\beta}{2} R_{\mu\nu} R^{\mu\nu}\right]\,,
\label{OldStelle}
\ee
where $\kappa_D^2=32\pi G_D$ is Newton's constant in $D$ dimensions and $\g$ and $\b$ are constants. A power-counting analysis gives the upper bound for the superficial degree of divergence for this theory: $\de= D L - 4 I + 4 V  = D - (D-4)(V-I) = D - (D-4)(V-I) = D + (D-4)(L-1)$, where $L$ is the number of loops, $V$ is the number of vertices and $I$ is the number of internal lines of the graph. We have substituted the topological relation $L = 1+ I -V$ in $\de$. In $D=4$ we get $\de=4$, so that all the divergences can be absorbed in the operators already present in the Lagrangian and the theory (\ref{OldStelle}) is renormalizable. It does contain an unstable spin-two mode, however, which makes the model nonunitary.

On the footprint of Stelle's theory, one can generalize \Eq{OldStelle} to a model renormalizable in any dimension \cite{Mod1}:
\ba
\fl {\mathcal L}_{D\text{-ren}} &=& a_1 R + a_2 R^2 + b_2 R_{\mu\nu}^2 + a_D R^{\frac{D}{2}} + b_D R_{\mu\nu}^{\frac{D}{2}}\nonumber\\
\fl &&+c_D R_{\mu\nu\rho\s}^{\frac{D}{2}} + d_D R \, \Box^{\frac{D}{2} - 2} R +\cdots  \,, \label{localDren} 
\ea
where covariant index contractions are implicit, $a_1=2/\kappa_D^2$, $a_2=\g/2$ and $b_2=\b/2$. This Lagrangian holds in even dimensions $D$; when $D$ is odd, it is sufficient to replace $D\to D-1$. Here, $\de=D$ and the theory is renormalizable.

The tree-level propagator for the action (\ref{OldStelle}) is
\ba
\fl \mathcal{O}^{-1} &=& \frac{\xi (2P^{(1)} + \bar{P}^{(0)} ) }{2 k^2 \, \omega_1( k^2)} 
+ \frac{P^{(2)}}{k^2 \Big(1 + \frac{k^2 \kappa_D^2 \beta}{4} \Big)}\nonumber\\
\fl &&- \frac{P^{(0)}}{2 k^2 \Big[\frac{D-2}{2}-k^2 \frac{D \beta \kappa_D^2/4+(D-1) \g \kappa_D^2}{2} \Big]}\,, \label{propagator}
\ea
where $\xi$ is a gauge parameter and $\omega_1(k^2)$ is a weight function \cite{Ste77}. The projectors are defined by $P^{(2)}_{\mu\nu\rho\s} = (\theta_{\mu\rho} \theta_{\nu\s} +\theta_{\mu\s} \theta_{\nu\rho})/2 - \theta_{\mu\nu} \theta_{\rho\s}/(D-1)$, $P^{(1)}_{\mu\nu\rho\s}= (\theta_{\mu\rho} \omega_{\nu\s} + \theta_{\mu\s} \omega_{\nu\rho} + \theta_{\nu\rho} \omega_{\mu\s}+ \theta_{\nu\s} \omega_{\mu\rho})/2$, $P^{(0)} _{\mu\nu\rho\s} = \theta_{\mu\nu} \theta_{\rho\s}/(D-1)$, $\bar{P}^{(0)} _{\mu\nu\rho\s} = \omega_{\mu\nu} \omega_{\rho\s}$, where $\theta_{\mu\nu} = \eta_{\mu\nu}-k_\mu k_\nu/k^2$ and $\omega_{\mu\nu} = k_\mu k_\nu/k^2$. 
Ignoring the tensorial structure, the quantum propagator at one loop in $D=4$ for the graviton is \cite{FrZ82}
\be
\mathcal{O}^{-1}_{2} \sim -\frac{1}{k^2\left[ 1 +\frac{\beta_2G_4}{2\pi} \,  k^2 \,  \ln \left( \frac{k^2}{E^2} \right)\right]}\,, \qquad \beta_2 = \frac{133}{10}\,,
\ee
where $E$ (usually denoted as $\mu$) is the renormalization scale. The constants in the form factor \Eq{ef} are
\be
a=\frac{133}{20\pi}\,G_4\,,\qquad n=1\,.
\ee
For the tachyon mode, the propagator is $\mathcal{O}^{-1}_{0} \sim -\{k^2[1- (3\beta_3G_4/\pi)k^2\times$ $\ln(k^2/E^2)]\}^{-1}$, where $\beta_3 \approx 0.101$. In this case, 
$a<0$ and the spectral-dimension analysis breaks down, as discussed above. We will ignore the tachyon and concentrate on the graviton.

In $D$ (even) dimensions, for the theory \Eq{localDren} we expect
\be
\mathcal{O}^{-1}_2
 \sim -\frac{1}{k^2\left[1+\left(c_0 k^2 +\dots+c_N k^{D-2}\right)\, \ln\left(\frac{k^2}{\mu^2}\right)\right]}\,, 
\ee
where the constants $c_i$ , as far as we know, have not been calculated yet. Assuming that $c_N>0$ for the graviton, we can consider the theory in even $D$ dimensions with inverse propagator given approximately by Eq.\ \Eq{ef} with
\be\label{anst}
a=c_N>0\,,\qquad n=\frac{D}{2}-1\,.
\ee

We now apply the results of the previous section. According to Eq.\ \Eq{anst}, the limit \Eq{dsuv} is
\be
\ds^{\rm UV}= 2\,,
\ee
independently of the number of topological dimensions. A two-dimensional UV regime is an almost universal feature of quantum-gravity models supposed to be UV finite or renormalizable. Stelle theory is one such model (even if it is nonunitary). 

\begin{figure}
\begin{center}
\includegraphics[width=8cm]{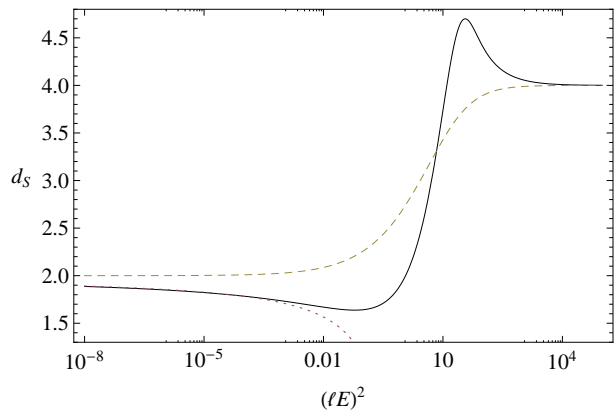}
\caption{\label{fig1} Solid curve: numerical evaluation of the spectral dimension \Eq{ds} with Eq.\ \Eq{ef} for $D=4$, $n=1$ and $\a=1$ (\emph{quantum} spectral dimension of Stelle gravity). Dotted curve: Eq.\ \Eq{dsuv} with the same parameter values. Dashed curve: spectral dimension for the polynomial dispersion relation \Eq{barf} for $D=4$, $n=1$ and $\a=1$ (\emph{classical} spectral dimension of Stelle gravity).}
\end{center}
\end{figure}

Figure \ref{fig1} shows the the full numerical evaluation of $\ds$ for $D=4$, $n=1$ and $\a=1$, together with the UV asymptotic behaviour \Eq{dsuv}. Contrary to the polynomial case \Eq{barf}, dimensional flow is not monotonic and displays a local minimum and a local maximum. These transient regimes appear only at the quantum level (the classical profile of $\ds$ is monotonic) and are very short in logarithmic scale.


\section*{Acknowledgments}
G.C.\ and L.M.\ acknowledge the i-Link cooperation program of CSIC (project ID i-Link0484) for partial sponsorship. The work of G.C.\ is under a Ram\'on y Cajal contract.



\appendix

\section{Massless limit}\label{A0} 

In the diffusion interpretation, one interprets the probe as a classical particle but, to justify Eq.\ \Eq{die}, the propagator \Eq{Gk} has been invoked. However, in a realistic treatment one should consider the quantum corrected propagator, take into account the physics of quantum fields and, in particular, all the difficulties related to the localization of massless particles (see Ref.\ \citen{Ska92} and references therein). By adding a nonzero mass, $\cK$ is deformed to a mass-dependent operator we shall denote as $\cK_m$. The dispersion relation changes to $\cK_m(k^2)+m^2=0$ and Eq.\ \Eq{Gk} becomes
\be\label{Gkm}
\tilde G(k^2)=-\frac{1}{\cK_m(k^2)+m^2}=-\int_0^{+\infty}\rmd(\ell^2)\,\rme^{-\ell^2 [\cK_m(k^2)+m^2]}\,.
\ee
If $m$ is the physical mass, then $\cK_m (-m^2)=-m^2$ (pole of the propagator at the physical mass) and $\cK_m'(-m^2)=1$ (unitarity; the prime is a derivative with respect to $k^2$), so that $\cK_m$ will have a branch point at  $-k^2 = M^2>m^2$, where $M$ is the mass production threshold of multi-particle states. Here and in the following, we shall always adopt a small-mass approximation for which $\cK_m (k^2)$ will be chosen to be independent of $m$ and equal to the form factor of the corresponding massless case, $\cK_m (k^2)  \simeq \cK (k^2)$, so that we shall omit the subscript $m$ in $\cK$.\footnote{In the on-shell renormalization scheme described above, the $m\to 0$ limit could be singular (IR catastrophe). In this limit, the minimal subtraction scheme should be adopted (then $m$ would not be the physical mass).} Then, $\cK(0)=0$ and all the mass dependence in the dispersion relation will be in the additive mass square term, 
\be\label{direl2}
\cK(k^2)+m^2=0\,.
\ee
The resulting diffusion equation and its solution are
\be\label{diem1}
\left[\frac{\p}{\p\ell^2}+\cK(-\N_x^2)+m^2\right]\,P_m(x,x';\ell)=0\,,\qquad P_m(x,x';0)=\de(x-x')\,,
\ee
where
\be
P_m(x,x';\ell)=\rme^{-\ell^2 m^2}P(x,x';\ell)\,.\label{diem2}
\ee
Upon replacing $\cP$ with $\cP_m$, Eqs.\ \Eq{fin1} and \Eq{P3} can serve as a consistency check of the small-mass approximation. In fact, from
\be
\cP_m(\ell)=\frac{\Om}{(2\pi)^D}\!\int_0^{+\infty}\rmd k\,k^{D-1}\,\rme^{-\ell^2 [\cK(k^2)+m^2]}=\rme^{-\ell^2 m^2}\cP(\ell)\,,\label{P2}
\ee
we can write $\ell^2 \cP_m(\ell) = -\p [ \cP_m(\ell)]/\p m^2$ that, substituted into \Eq{fin1} and compared with \Eq{P3}, leads to  
\be
\int\rmd^Dx\,|G(x-x')|^2 =- \frac{\p}{\p m^2} \int_0^{+\infty}\rmd(\ell^2)\,\cP_m(\ell) = \frac{\p}{\p m^2} G(0)\,.\label{tst1}
\ee
This is indeed satisfied in the small-mass approximation, as it is equivalent to the identity
\be
\frac{1}{[\cK(k^2) + m^2]^2}=-\frac{\p}{\p m^2} \ \frac{1}{\cK(k^2) + m^2}\,.
\ee

Looking at Eq.\ \Eq{diem2}, we see that all the mass dependence is in an overall exponential factor, which 
 does not modify the initial condition but it changes the normalization \Eq{usno}. From Eq.\ \Eq{usno},
\be\label{usnom}
\int\rmd^Dx\,P_m=\rme^{-\ell^2 m^2}\leq 1\,.
\ee
The interpretation of $P_m$ as a probability density is in accordance with the fact, well known in transport theory, that the mass is a dissipative term. Therefore, the probability is not conserved along $\ell$. In the diffusion-probabilistic interpretation of $\ds$, the total probability is imposed to be 1: we must find the probe somewhere in the inspected volume with almost certainty. This is just an operational definition corresponding to take the limit $m\to 0$ and is necessary to recover the correct value of $\ds$ even in the simplest classical case (see Eq.\ \Eq{dsm}). However, if one gives up the statistical interpretation of the spectral dimension and adopts a QFT interpretation, sending the mass to zero may be questionable, not only because the most generic probe will possess a mass but also because, strictly speaking, a massless scalar particle cannot be localized. This means that, from a relativistic point of view, a pointwise particle with $m=0$ may not make much physical sense as an initial condition. Therefore, the problem we pose ourselves is how to interpret the massless limit in the QFT setting.

Having $m\neq 0$ makes the resolution interpretation proposed in the text more credible. If we probed the geometry with bad enough resolution, it should become possible (i.e.\ with nonzero probability) to miss the test particle in a measurement and see it nowhere. This is precisely the meaning of the normalization \Eq{usnom}: as the resolution worsens with respect to the mass scale (large $\ell m$), the probability to find the particle ``somewhere'' on the geometry diminishes. Note also that the parameter $m$ in \Eq{Gkm} is to be identified with the physical mass of the field, not with the bare mass $m_0$ appearing in the classical Lagrangian. This is because we assumed the dispersion relation \Eq{direl2}, which is the definition of physical mass \cite{Sre07}. Therefore, $m$ is a constant independent of the resolution $1/\ell$ and we treated it as such in Eqs.\ \Eq{diem1} and \Eq{diem2}.

Let us now discuss the effect of $m$ on the spectral dimension. Even when $P_m(x,x';\ell)$ is not positive definite, its trace indeed is (such is the reason why the negative-probabilities problem has been overlooked in early literature). This is all what is needed to interpret $\cP_m(\ell)$ as the return probability (at a given resolution scale $\ell$) and to evaluate the spectral dimension. 

From Eq.\ \Eq{P2} and from the positivity of $\cP$ it immediately follows that, as anticipated, $\cP_m(\ell)\geq 0$ for all $\ell$. If we used $\cP_m$ instead of $\cP$ in the definition of the spectral dimension, we would get the quantity
\be\label{dsm}
\ds^{(m)}(\ell):=-2\frac{\p\ln\cP_m(\ell)}{\p\ln\ell^2}=\ds(\ell)+2\ell^2m^2\,,
\ee
to be compared with \Eq{ds}. The mass does not modify the spectral dimension in the limit of infinite resolution $\ell\to 0$ (UV limit), but it produces a divergence in the IR limit (or maximum coarse graining) $\ell\to +\infty$. For geometries without dimensional flow, the correct value of the spectral dimension is obtained by asking to take the infinite-resolution limit, i.e.\ $\ds=\lim_{\ell\to 0}\ds^{(m)}$. In all other cases, we can simply define the spectral dimension as Eq.\ \Eq{ds}.

We can see that the limit $m\to 0$ is indeed a sensible requirement by considering the simple example of ordinary Euclidean space, Eq.\ \Eq{fk}. There, $\ds^{(m)}(\ell)=D+2\ell^2m^2$ ceases to be a viable geometric indicator at energy scales $\ell^{-1}\lesssim m$ lower than the particle mass, where $\ds^{(m)}$ blows up for coarse resolution (large $\ell$). For the test particle to be an efficient probe of geometry, one must excite it with energies greater than its mass, i.e.\ $\ell m\ll 1$ for any $\ell$. Consequently, very massive particles are not good probes since they establish an upper limit to the allowed resolution of order $\sim 1/m$. We conclude that, contrary to Eq.\ \Eq{dsm}, Eq.\ \Eq{ds} does not contain spurious effects due to the characteristics of the particle. For all purposes, in the main text we ignore the mass, bearing however in mind that the tension between this prescription and the QFT requirement of having a massive localizable probe has been easily addressed by implementing the resolution interpretation.


\section{Small-resolution (IR) limit}\label{A1}

In this section, we study the integral \Eq{P2} with the form factor \Eq{ef} (with $a>0$) in the limit of small resolution $\ell^{-1}\ll E$. Calling $y:=k^2/E^2$ and $\cI:=2\cP (2\pi)^D/(\Om E^D)$, we have to estimate
\be\label{Py}
\cI=\int_0^{+\infty}\rmd y\,y^{\frac{D}{2}-1}\,\rme^{-\s \cK(y)}\,,
\ee
where  $\s:=(\ell E)^2$ is dimensionless,
\be\label{fofa}
\cK(y)=y(1+\a\, y^n\ln y)
\ee
is the form factor and $\a=a E^{2n}>0$ is a dimensionless constant. A special function we will often use is the Euler function (formula 8.310.1 of Ref.\ \citen{GR})
\be\label{eu}
\G(c):=\int_0^{+\infty}\rmd z\,z^{c-1}\,\rme^{-z}\,.
\ee
At first, one might try to compute \Eq{Py} by expanding part of the exponential while retaining a dumping factor, and then assuming that the sum and integration operations commute:
\be\label{boh}
\cI=\sum_{m=0}^{+\infty}\frac{(\a\s)^m}{m!}\int_0^{+\infty}\rmd y\,y^{\frac{D}{2}-1+m(n+1)}\,\rme^{-\s y}(-\ln y)^m.
\ee
Using iteratively Eq.\ \Eq{eu} and some summation manipulation, one ends up with ($q=m-l$)
\be
\fl\cI =\frac{1}{\s^{\frac{D}{2}}}\sum_{l=0}^{+\infty}\sum_{q=0}^{+\infty}\frac{(-2)^l}{l!q!}\left(\frac{\a}{\s^n}\right)^{l+q}(\ln\s)^q\frac{\p^l}{\p D^l}\G\left[\frac{D}{2}+(n+1)(l+q)\right].\label{sum}
\ee
A similar expression is provided also by formula 4.358.5 of Ref.\ \citen{GR}. 

The small-resolution (IR) limit corresponds to
\be
\s\gg 1\,,
\ee
for which the $(l,q)=(0,0)$ term in the sum \Eq{sum} dominates. Then, $\cP\sim \s^{-D/2}$ and, from Eq.\ \Eq{ds}, one gets the expected result $\ds\simeq D$, Eq.\ \Eq{dsir}. 
However, in commuting the sum with the integral in Eq.\ \Eq{boh} we have thrown away vital information for the correct evaluation of $\cI$. To recover it, we restart the analysis from the study of the form factor \Eq{fofa}. 

The first derivative of $\cK$ with respect to $y$ is $\cK'(y)=1+\a y^{n}[1+(n+1)\ln y]$, which vanishes at
\be\label{yn}
\bar y^{(i)}:=\left\{-\frac{n}{\a(n+1)\,W_i\left[-\frac{n\rme^{n/(n+1)}}{\a(n+1)}\right]}\right\}^{\frac1n}.
\ee
\begin{figure}
\begin{center}
\includegraphics[width=8cm]{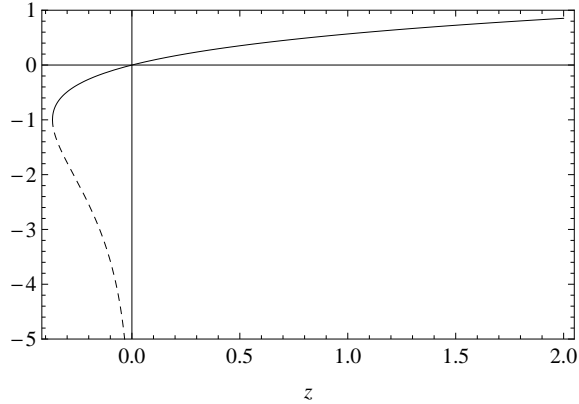}
\end{center}
\caption{\label{fig2} The two real branches of the Lambert function $W_{-1}(z)$ (dashed curved) and $W_0(z)$ (solid curve).} 
\end{figure}
Here $W_i$ are the two real branches of Lambert's function $W$ drawn in Fig.\ \ref{fig2}, defined implicitly by the transcendental equation (Sec.\ 4.13 of Ref.\ \citen{NIST})
\be\label{lam}
z=W(z)\,\rme^{W(z)}\,.
\ee
This equation has real solutions only if $z\geq-1/\rme$, corresponding, in Eq.\ \Eq{yn}, to $\a\geq \bar\a$, where
\be
\bar\a:=\frac{n}{n+1}\,\rme^{(2n+1)/(n+1)}.
\ee
In this case, there are two real branches of $W$, denoted by $W_i$. One, called $W_{-1}$, is defined in the interval $-1/\rme<z\leq 0$ and runs from $W_{-1}(0)=-\infty$ to $W_{-1}(-1/\rme)=-1$; the other, called $W_{0}$, is defined in the interval $z>-1/\rme$ and runs from $W_{0}(-1/\rme)=-1$ to infinity, with $W_0(0)=0$ (see Fig.\ \ref{fig2}). In Ref.\ \citen{NIST}, $W_0$ and $W_{-1}$ are denoted by $W_p$ and $W_m$, respectively.

In the case $\a>0$, we are only interested in the semi-interval $z<0$, where we have two roots of the form \Eq{yn} with $W_{-1}$ and $W_0$, respectively corresponding to a local maximum and a local minimum of $\cK$. In the special case $n=0$, there is only one root
\be\label{n0}
y_0=\rme^{-1-1/\a}\,,\qquad \bar\a=0\,,
\ee
corresponding to an absolute (negative) minimum, $\cK(y_0)=-\a y_0$. If $\a<0$, the only root (\Eq{yn} with $W_0$ if $n\neq 0$ or \Eq{n0} for $n=0$) corresponds to an absolute positive maximum.

\begin{figure}
\begin{center}
\includegraphics[width=8cm]{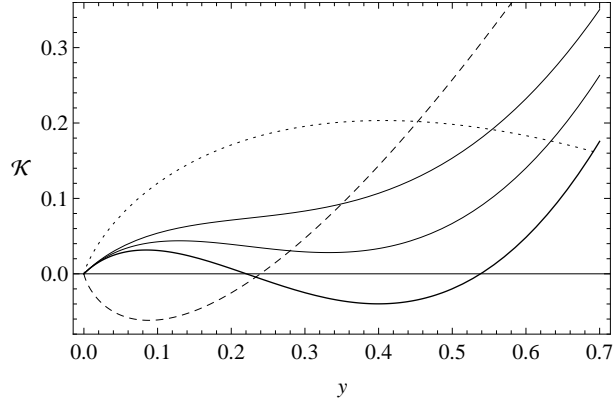}
\end{center}
\caption{\label{fig3} Solid curves: the form factor \Eq{fofa} with $n=1$ ($\bar\a=\rme^{3/2}/2$, $\a_{*}=\rme$) and, with increasing thickness, $\a=2,2.5,3$, corresponding to cases (a)--(c). Dashed curve: $n=0$ (case (d)) with $\a=0.7$. Dotted curve: case (e) with $n=0$, $\a=-10$ and $\cK$ rescaled by $1/20$ to make the maximum enter the range of the plot.}
\end{figure}

Denote with $\bar y$ the position of the minimum, that is $\bar y^{(i=0)}$. We omit the index $i=0$ to keep notation light. Another critical point in the parameter space occurs when the local minimum of $\cK$ changes sign. This happens when $\cK(\bar y)=\bar y(1+\a \bar y^n\ln \bar y)=0$ but, since $\bar y\neq 0$, this reduces to $0=1+\a \bar y^n\ln \bar y=(n-\a \bar y^n)/(n+1)$, where we used the condition $\cK'(\bar y)=0$. Therefore, $\cK(\bar y)=0$ for $\bar y=(n/\a)^{1/n}$. From Eq.~\Eq{yn} and calling $z_*$ the argument of the Lambert function therein, it follows that $W(z_*)=-1/(n+1)$. Plugging this into Eq.\ \Eq{lam}, one finally gets the critical value $\a_{*}=n\,\rme\geq \bar\a$, where the equality holds only if $n=0$. Thus, we recognize five cases (Fig.\ \ref{fig3}):
\begin{enumerate}
\item[(a)] $0<\a<\bar\a$, $n\neq 0$: $\cK'(y)>0$ for all $y$ and $\cK>0$ is monotonic.
\item[(b)] $\bar\a<\a<\a_{*}$, $n\neq 0$: $\cK$ has a local maximum followed by a local minimum, both positive.
\item[(c)] $\a>\a_*$, $n\neq 0$: $\cK$ has a positive local maximum followed by a negative absolute minimum.
\item[(d)] $\a>\bar\a=\a_{*}=0$, $n=0$: $\cK$ has one negative absolute minimum.
\item[(e)] $\a<0$, any $n$: $\cK$ has one positive absolute maximum.
\end{enumerate}
In the critical cases $\a=\bar\a$ and $\a=\a_{*}$, the second extremum is, respectively, a saddle point and a minimum at $\cK=0$.

Let us now calculate the integral \Eq{Py}, beginning with case (a). Here, the large-$\s$ limit is easily obtained by expanding the exponential $\exp(-\s \a y^n\ln y)=1-\s\a y^n\ln y+\dots$ and retaining only the first term; the others are subleading, since $n>0$. Using the definition \Eq{eu}, we obtain
\be
\fl\cI_{{(a)}} =\int_0^{+\infty}\rmd y\,y^{\frac{D}{2}-1}\,\rme^{-\s y}\left(1-\s\a y^n\ln y+\cdots\right) =\frac{\G\left(\frac{D}{2}\right)}{\s^{\frac{D}{2}}}\left[1+O\left(\frac{\ln\s}{\s^n}\right)\right],\label{Pya}
\ee
where the order of the first next-to-leading term can be checked using formula 4.352.1 of Ref.\ \citen{GR}. 

To deal with case (b), we split the integration range $[0,+\infty)$ at the position of the local maximum $y_{\rm max}=\bar y^{(-1)}$, $[0,y_{\rm max}]\cup [y_{\rm max},+\infty)$. Up to $y_{\rm max}$, the integral \Eq{Py} is approximated with $\cI_{{(a)}}$, while the integral between $y_{\rm max}$ and infinity can be estimated with the Laplace method \cite{copson65}. Let $f(y)$ be a twice-differentiable function with a single minimum in the interval $(A,B)$, with $A$ and $B$ possibly infinite, and let $g(y)$ positive and bounded in the same interval. The Laplace approximation states that, in the large-$\s$ limit, the main contribution to the integral $\int_A^B\rmd y\, g(y) \, \rme^{-\s f(y)}$ comes from points near the minimum. One expands $f$ and $g$ around the minimum, $f(y)=f(y_{\rm min})+(1/2) f''(y_{\rm min})(y-y_{\rm min})^2+\cdots$, $g(y) = g(y_{\rm min}) +\cdots$, so that
\ba
\int_A^B\rmd y\, g(y) \, \rme^{-\s f(y)} &\simeq & g(y_{\rm min})\, \rme^{-\s f(y_{\rm min})}\int_{-\infty}^{+\infty}\rmd y\,\rme^{-\frac12\s f''(y_{\rm min})(y-y_{\rm min})^2}\nonumber\\
&=&g(y_{\rm min})\, \rme^{-\s f(y_{\rm min})}\sqrt{\frac{2\pi}{\s f''(y_{\rm min})}}\,.
\ea
In the present case, $f(y)=\cK(y)$, $g(y)=y^{(D/2)-1}$ and $y_{\rm min}=\bar y^{(0)}$, which yields
\be
\int_{y_{\rm max}}^{+\infty}\rmd y\,y^{\frac{D}{2}-1}\,\rme^{-\s \cK(y)} \simeq y_{\rm min}^{\frac{D}{2}-1}\rme^{-\s \cK(y_{\rm min})}\sqrt{\frac{2\pi}{\s \cK''(y_{\rm min})}} =: \tilde\cI\,.\label{tiI}
\ee
Overall,
\be\label{Pyb}
\cI_{{(b)}} = \cI_{{(a)}} +\tilde\cI \simeq  \cI_{{(a)}}\,,
\ee
since $\cK(y_{\rm min})>0$ as long as $\a<\a_{*}$ and $\tilde\cI$ is exponentially suppressed. On the other hand, both in case (c) and (d) ($n=0$) $\cK(y_{\rm min})<0$, so that $\tilde\cI$ dominates:
\be\label{Pycd}
\cI_{{(c)}} = \cI_{{(a)}} +\tilde\cI \simeq  \tilde\cI\simeq \cI_{{(d)}}\,.
\ee
Finally, in case (e) ($\a<0$) the integral diverges for any $\s$.


\section{High-resolution (UV) limit}\label{A2}

To evaluate the integral \Eq{Py} in the high-resolution limit
\be
\s\ll 1\,,
\ee
we change variable $x:= y\s^{1/(n+1)}$ and get
\be\label{Pyuv}
\cI=\s^{-\frac{D}{2(n+1)}}\int_0^{+\infty}\rmd x\,x^{\frac{D}{2}-1}\,\rme^{-\frac{\a}{n+1}(-\ln\s)x^{n+1}}\rme^{-x\s^{n/(n+1)}-\a\,x^{n+1}\ln x}\,.
\ee
The second exponential is bounded from above and below for all $\s>0$, $\a$ and $x>0$, and it can be expanded as $1-x\s^{n/(n+1)}-\a\,x^{n+1}\ln x+\cdots$.
The first exponential dominates when $\s$ is small, in a neighborhood of $x=0$ for $\a>0$. Using again Eq.~\Eq{eu},
\be\label{ciuv}
\cI\ \stackrel{\s\ll 1}{=} \frac{\G\left[\frac{D}{2(n+1)}\right]}{(n+1)\,\left(-\frac{\a}{n+1}\s\ln\s\right)^{\frac{D}{2(n+1)}}}\left[1+O\left(\frac{1}{\ln\s}\right)\right].
\ee



\end{document}